\documentclass[PAPER, atlasdraft=false, coverpage=false, cernpreprint, USenglish, texmf, orcidlogo]{atlasdoc}
\usepackage{atlaspackage}
\usepackage{atlasbiblatex}

\usepackage{atlasphysics}
\usepackage{atlasheavyion}
\graphicspath{{logos/}{figures/}}

\usepackage{ANA-HION-2024-12-PAPER-defs}

\AtlasTitle{Observation of impact parameter dependent modifications of
nuclear parton distributions
in photonuclear \ce{Pb}+\ce{Pb} collisions at $\sqrt{s_{_\text{NN}}} = 5.02$ TeV with the ATLAS detector}

\AtlasAbstract{%
High-energy photonuclear ($\gamma+A$) scattering in ultra-peripheral
heavy-ion collisions provides a unique probe of nuclear structure.
This Letter studies the dependence of $\gamma+A$ jet production in ultra-peripheral
\ce{Pb}+\ce{Pb} collisions at $\sqrt{s_{_\text{NN}}} = 5.02$ TeV on the presence of forward neutron
emission from either nucleus. The data was taken in 2018 with the ATLAS
detector at the LHC and corresponds to an integrated luminosity of
$1.72$ nb$^{-1}$. The kinematics of the hard $\gamma+A$ processes,
expressed via the particle-level photon ($z_{-}$) or nuclear parton
($x_{+}$) momentum fractions, are determined from $R = 0.4$ jets
reconstructed using the anti-$k_t$ algorithm.
At lower $z_{-}$, where
the non-diffractive component dominates, the nuclear parton
distribution can be cleanly probed in collisions that leave
the struck nucleus essentially intact. Such collisions are expected to
probe larger impact parameters ($b_\text{A}$) within the target.
The shape of the $\gamma+A$ cross-section as a function of $x_{+}$ in such
collisions is found to differ from that in $\gamma+A$ collisions accompanied
by forward neutron emission, with an observed significance of $6.0\sigma$.
These results are consistent at large $x_{+}$ with large $b_\text{A}$ collisions
exhibiting no modifications to the parton distributions that are
usually observed in hard scattering processes involving nuclei, relative to
collisions with smaller $b_\text{A}$.
Thus, these measurements provide an experimental observation that the
modifications to nuclear parton distributions vary with impact parameter.
}

\AtlasRefCode{HION-2024-12}

\PreprintIdNumber{CERN-EP-2026-095}

\AtlasJournal{Phys.\ Rev.\ Lett.}
\AtlasCoverEgroupAnalysisTeam{atlas-HION-2024-12-analysis-team@cern.ch}

\hypersetup{pdftitle={ATLAS document},pdfauthor={The ATLAS Collaboration}}

\begin{document}

\maketitle

%

%
%

%
%
%

In recent years, there has been significant interest in nuclear parton
distribution functions (nPDFs) and their modifications relative to free nucleon
distributions. This interest is driven in part by the heavy-ion programs at the
Relativistic Heavy Ion Collider and Large Hadron Collider, in which the study
of hard-scattering processes is a crucial component. The possibility
that nPDF modifications may depend on the impact parameter
relative to the center of the struck nucleus ($b_\mathrm{A}$)
has long been considered~\cite{Burkardt:2002hr,Helenius:2012wd},
although nearly all existing fits neglect any such dependence.
If present, this effect would imply that peripheral (large $b_\mathrm{A}$)
collisions probe different nPDFs than inclusive collisions.

Descriptions of the suppression of nPDFs at low $x$ (shadowing) typically
involve multiple scattering, which would necessarily imply a dependence on $b_\mathrm{A}$
\cite{Guzey:2017sz}. Also, recent interpretations of the suppressed nPDF at
larger $x$, the ``EMC'' effect~\cite{EuropeanMuon:1983wih}, in terms of
short-range correlations (SRCs)~\cite{Hen:2012fm,Hen:2016kwk} might
suggest a dependence on $b_\mathrm{A}$ due to variations in the nuclear
density and presence of a neutron skin in large nuclei~\cite{Huth:2021bsp}.
The $b_\mathrm{A}$-dependent nPDFs are also related to
the Fourier transform of a special case of the nuclear Generalized
Parton Distributions (nGPDs)~\cite{Burkardt:2002hr}. These
distributions connect inclusive measurements of nPDFs to
exclusive vector meson photoproduction results~\cite{CMS-HIN-22-002,ALICE:2024whv,STAR:2023nos}
that probe the nGPD~\cite{Eskola:2024fhf}.
While several experimental probes of the  $b_\mathrm{A}$-dependence of nPDFs have been
proposed~\cite{Klein:2003dj,PHENIX:2010hmo,McGlinchey:2012bp,Shao:2020acd}, none have yet
succeeded in demonstrating or directly constraining this dependence.

\begin{figure}[!b]
\centerline{
\subfloat[]{\includegraphics[width=0.425\textwidth]{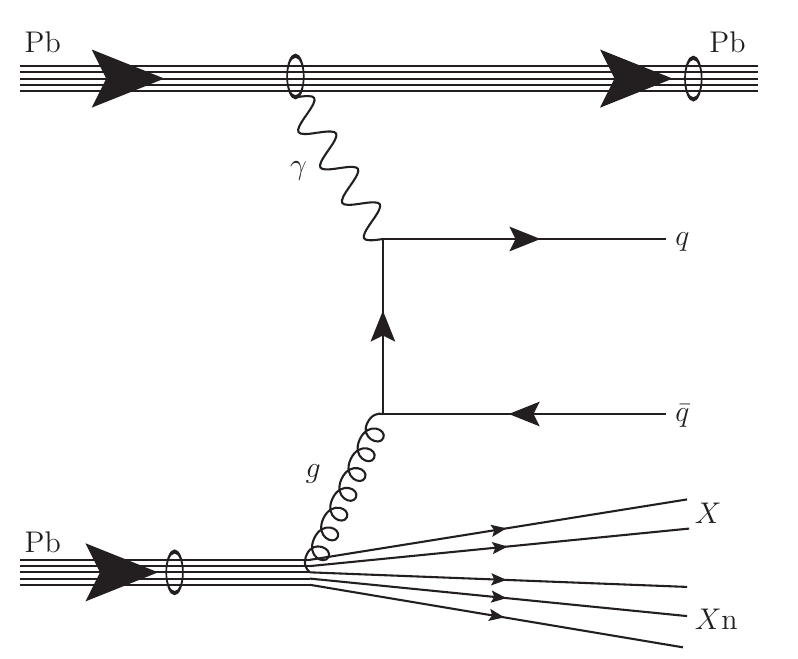}}
\subfloat[]{\includegraphics[width=0.465\textwidth]{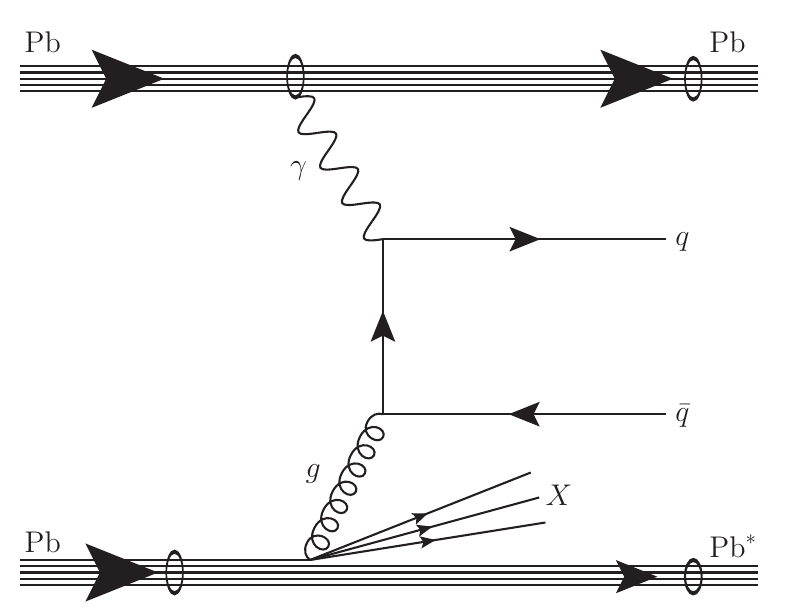}}
}
\caption{
Feynman diagrams describing the direct photon case for the two processes
measured in this Letter. (a) A \gamA\ interaction where the photon strikes
a nucleon near the center of the nucleus, producing a pair of jets while
breaking up the nucleus and emitting forward neutrons. (b) A \gamA\ interaction
where the photon strikes a nucleon near the edge of the nucleus, producing a pair
of jets while leaving the rest of the nucleus intact. In both diagrams, $X$ denotes
additional particle production due to color connections between the beam and the
jet system on the side of the struck nucleus.
}
\label{fig:Feynman}
\end{figure}

ATLAS has performed several measurements~\cite{HION-2018-27,HION-2024-02,HION-2022-15}
of inclusive photonuclear (\gamA) scattering in
ultra-peripheral collisions (UPCs), where the nucleus-nucleus
impact parameter ($b_\mathrm{AA}$) is greater than twice the nuclear
radius, so that only
photon-induced processes contribute.
A recent ATLAS measurement~\cite{HION-2022-15} studied \gamAjets\
production in UPCs, where
events are characterized by the number of neutrons
emitted from each nucleus, with $Xn$ implying at least one neutron and
$0n$ indicating no emitted neutrons.
Photonuclear collisions typically have a \znxn\ topology (panel (a) of Figure~\ref{fig:Feynman}),
as the photons are usually emitted coherently by the source nucleus,
which is left un-excited~\cite{Klein:2016yzr}. Conversely, the hard
scattering process has a very high probability to excite the struck
nucleus, causing the emission of one or more neutrons.

The $b_\mathrm{A}$ dependence of \gamA processes can be
characterized using the number of neutrons emitted from the struck
nucleus. The relationship between $b_\mathrm{A}$ and the number of emitted neutrons
depends on both the physics of the scattering process and on the
de-excitation cascade of the resulting excited
nucleus~\cite{Strikman:1998cc,Larionov:2018igy,Alvioli:2024cmd}.
However, on general grounds, a lower number of emitted neutrons suggests photonuclear collisions at larger
$b_\mathrm{A}$. Thus, the class of collisions with no forward
neutrons (panel (b) of Figure~\ref{fig:Feynman}) would be expected to probe large $b_\mathrm{A}$.
Such collisions, having a \znzn\ topology, necessarily involve
either a very peripheral collision or one
that does not sufficiently excite the struck nucleus to induce emission
of recoil or evaporation neutrons.
A potential background to the photonuclear process arises from
diffractive photonuclear processes involving coherent pomeron ($I\!\!P$) emission by the
struck nucleus (\gamPom) that leaves it
intact~\cite{Guzey:2016tek}. These processes are interesting in
their own right and have been studied comprehensively in $\gamma+p$
collisions at HERA~\cite{ZEUS:2007uvk,H1:2010xdi,H1:2012xlc,H1:2015okx}.

Photonuclear hard-scattering processes in UPCs may also be accompanied
by additional soft photon exchanges which cause electromagnetic
dissociation (EMD) and subsequent neutron emission from one or both
nuclei in a direction un-correlated with any features of the central
event~\cite{PhysRevLett.109.252302,ALICE:2022iqi}. The probability for additional EMD
increases with decreasing $b_\textrm{AA}$~\cite{CMS-HIN-19-014,CMS-HIN-22-002,Guzey:2013jaa}
so that the presence or absence of ``extra'' neutrons, for example in the direction of
the photon-emitting nucleus in \gamA, provides indirect sensitivity to $b_\mathrm{A}$
(see Ref.~\cite{Eskola:2024fhf} for a quantitative assessment). Measurements
presented in Ref.~\cite{OtherPaperPRC} indicate that the rate for additional EMD of
the photon-emitting nucleus in
\znzn\ collisions is lower than that in \znxn\ collisions. This observation provides
additional, independent evidence supporting the claim that \znzn\ \gamA\ collisions
probe larger $b_\mathrm{A}$ than \znxn\ collisions.

This Letter presents measurements of \gamAjets\ scattering in
\PbPb\ UPCs at $\sqn = 5.02$~\TeV with no
forward neutrons, and thus, a \znzn\ topology. They are compared to additional
measurements of \gamAjets\ processes in \znxn\ events with otherwise identical
selections. At forward jet rapidity, the non-diffractive contribution to the
\znzn\ cross-section can be cleanly selected using rapidity gap requirements in a
kinematic region where the diffractive contribution is suppressed. In
this region, the ratio of the \znzn\ and \znxn\ cross-sections is
measured, providing direct sensitivity to the impact parameter dependence of nPDF
modifications, relative to free nucleons.


%

%
%
%

The measurements presented here were performed using the ATLAS
calorimeter, inner detector (ID), trigger, and data acquisition
systems~\cite{PERF-2007-01}. The calorimeter system consists of a
liquid argon (LAr) electromagnetic (EM) calorimeter ($|\eta|<3.2$), a
steel/scintillator sampling hadronic calorimeter ($|\eta| < 1.7$), a
LAr hadronic calorimeter ($1.5 < |\eta| < 3.2$), and a forward
calorimeter (FCal) ($3.2 < |\eta| < 4.9$).  The inner tracking
detector covers the pseudorapidity range \(|\eta| < 2.5\) and consists
of silicon pixel, silicon microstrip, and transition radiation
tracking detectors all immersed in a 2~T axial magnetic field.\footnote{ATLAS uses a right-handed coordinate system with its origin at the nominal interaction point (IP)
in the center of the detector and the \(z\)-axis along the beam pipe.
The \(x\)-axis points from the IP to the center of the LHC ring,
and the \(y\)-axis points upwards.
Polar coordinates \((r,\phi)\) are used in the transverse plane,
\(\phi\) being the azimuthal angle around the \(z\)-axis.
The pseudorapidity is defined in terms of the polar angle \(\theta\) as \(\eta = -\ln \tan(\theta/2)\) and is equal to the rapidity
$ y = \frac{1}{2} \ln \left( \frac{E + p_z c}{E - p_z c} \right) $ in the relativistic limit.
Angular distance is measured in units of \(\Delta R \equiv \sqrt{(\Delta y)^{2} + (\Delta\phi)^{2}}\).}

Two zero-degree calorimeters~\cite{Avoni:2025lmu} (ZDCs) measure neutrons emitted at small rapidity
separation from the incident nuclei. For the detector configuration
relevant to this measurement, they are tungsten-quartz sampling
calorimeters with a total of 4.4 interaction lengths of absorber that
are located symmetrically at a distance of
$\pm 140$~m from the IP and that cover $|\eta|>8.3$.
The luminosity is measured mainly by the LUCID--2~\cite{LUCID2} detector.

A two-level trigger system is used to select
events~\cite{TRIG-2016-01,TRIG-2019-04}. The first-level (L1) trigger is implemented in hardware and uses a
subset of the detector information to accept events at a rate close to \qty{100}{\kHz}. Both
calorimeter and ZDC inputs may be used to accept events in the L1 trigger.
Fully assembled events passing the L1 trigger are analyzed and selected for recording using software-based algorithms in the  ``High-Level'' trigger (HLT). A software suite~\cite{SOFT-2022-02} is used in data simulation, in the reconstruction
and analysis of real and simulated data, in detector operations, and in the trigger and data acquisition
systems of the experiment.


%

%
%

%
%
%

These measurements use $1.72$~nb$^{-1}$ of
\PbPb\ collision data at $\sqn = 5.02$~\TeV recorded by ATLAS in 2018.
The primary triggers used for the measurement involve a combination of
\znzn\ (double veto) ZDC requirements and total transverse energy
requirements in the calorimeter at
L1, as well as a requirement of at least one jet above a given \pT threshold
in the HLT. The L1 triggers require a total transverse energy (\sumET) in the calorimeter
which satisfies $5 < \sumET < 200$~\GeV.
The HLT jet triggers are based on the anti-\kt
algorithm~\cite{Cacciari:2008gp,Fastjet} with radius $R = 0.4$ applied to
topological clusters (topo-clusters)~\cite{PERF-2014-07}
formed from energy deposits in the calorimeter. The lowest \pT
trigger threshold used in this measurement was fully %
efficient for single jets at 13~\GeV (20~\GeV) for $|\eta| > 3.2$
($|\eta|<3.2$). The efficiency for the HLT trigger to select any
of the jets in the final state is greater than 98\% for all events within
the fiducial acceptance of the measurement.
To collect a sample of \znxn\ jet events, a separate set of triggers were deployed
using identical jet and \sumET\ requirements but with an L1 \znxn\ ZDC trigger.

Monte Carlo (MC) simulated samples were produced using
the \pythia\ event generator~\cite{Bierlich:2022pfr,Helenius:2019gbd} with the A14
set of tuned parameters~\cite{ATL-PHYS-PUB-2014-021} for three relevant
physical processes: photonuclear (\gamAjets), photon--pomeron (\gamPomJets),
and photon--photon (\gamGamJets). The photon flux was
computed using \textsc{Starlight}~\cite{Klein:2016yzr}, with the integration over impact
parameter and the target density performed using methods described in Ref.~\cite{HION-2022-15}.
For the \gamAjets\ process, the samples were produced using nCTEQ15~\cite{Kovarik:2015cma}
nuclear PDFs. Since \pythia\ does not simulate forward neutron emission, the same simulated
sample is used to model both the \znzn\ and \znxn\ \gamAjets\ events.
Hard diffraction with a user-specified pomeron ($I\!\!P$) flux was added in \pythia version
8.308 to simulate the process with coherent nuclear pomeron emission. For each of
the three samples, the simulated signal events include
both direct and resolved photon processes; the latter require additional
modeling using the CJKL photon PDF set~\cite{Cor03}.
Final-state stable particles, defined as those with $c\tau > 10$~mm, were then
passed to a \textsc{Geant4}-based simulation of the ATLAS detector
\cite{Agostinelli:2002hh,SOFT-2010-01}, the output of which was
reconstructed in the same way as data.

Particle-Flow~\cite{PERF-2015-09} jets are reconstructed using the implementation of
the anti-$k_t$ algorithm~\cite{Cacciari:2008gp} with radius parameter $R = 0.4$ in
the \verb+FastJet+~\cite{Fastjet} package. These jets use combined calorimeter and
tracking information, and a dedicated calibration designed for detector conditions in
UPC~\cite{HION-2022-15} is applied, derived using methods described in Ref.~\cite{JETM-2018-05}.
Every event must have at least two reconstructed jets
with the fiducial requirements: $\pTjet > 13$~\GeV and $\etajet < 4.4$.
From jets satisfying these requirements, the mass (\mjets), rapidity (\yjets), and
total scalar transverse momentum ($\HT \equiv \sum_i
p_{\mathrm{T}\,i}^\mathrm{jet}$), of the $N$-jet system resulting from the hard-scattering
process are computed. A requirement is also imposed of $0.9 < \mjets/\HT < 4$,
consistent with Ref.~\cite{HION-2022-15}, to reduce the impact of jets not
originating from the primary hard-scattering.

Offline requirements are applied to select events recorded during stable
running conditions of the LHC, that have no detector hardware or readout
error, and that are not consistent with beam-induced backgrounds~\cite{HION-2022-15}.
UPC \gamA events are selected
using a combination of offline jet, ZDC, and rapidity gap requirements, in addition
to jet selections. Identical event selections (aside from ZDC requirements) are
applied to the \znzn\ and \znxn\ samples.
The \znzn\ requirement is applied through a veto on events
having energy greater than 40\% of the beam energy, i.e. more than
1~\TeV, in either ZDC.
The sum of rapidity gaps on either side of
the jets ($\sumGap$) is computed using the methodology from Ref.~\cite{HION-2022-15},
and the maximum sum of rapidity gaps ($\sumGapMax$) is taken as the larger of the two values.
To exclude hadronic \PbPb\ collisions, this quantity is required to
satisfy $\sumGapMax > 2.5$. This requirement removes $22.9\%$
of \znzn\ events passing all other selections.

The non-diffractive \gamA collisions are intrinsically asymmetric, as the photon energies
are usually much smaller than the longitudinal momenta of the struck partons. Thus,
the sign of \yjets is assumed to indicate the direction of the struck nucleus.
Using this assumption, which according to the \gamAjets\ MC sample is
valid for more than $99.9\%$ of the studied events, two momentum
fractions are defined as:
\begin{align}
\zneg \equiv\frac{ \mjets }{\sqn} e^{-|\yjets|},~ \xpos \equiv\frac{ \mjets}{\sqn} e^{+|\yjets|}.
\end{align}
Neglecting initial and final-state radiation and beyond-LO
contributions, these correspond to fractions of the beam momentum carried by the
partons entering the hard-scattering process from the emitted photon (\zneg) and
struck nucleus (\xpos).

Event-by-event corrections are applied to the data to account for trigger inefficiencies,
which are computed for events that pass all other event selections.
Possible trigger inefficiency arises from three sources:
the L1 ZDC selections, L1 \sumET\ thresholds, and HLT jet requirements. Studies of
the L1 ZDC trigger efficiency indicate that it is
greater than $99.9\%$ efficient for selecting \znzn\ and
\znxn\ events. The efficiency of the L1 \sumET\ and jet requirements
are found to rise rapidly at low \HT, such that both have at
most a $5\%$ inefficiency for  $\HT > 28$~\GeV, for which a correction is applied.
The veto on events with $\sumET > 200$~\GeV\ is found to remove no events for the
jet kinematics studied, $\HT < 150$~\GeV. Both the \sumET\
and jet trigger inefficiencies affect the \znzn\ and \znxn\ samples in the same way,
so their impact on the ratio is negligible. The efficiency of the $\sumGapMax > 2.5$
requirement is  assessed using \pythia\ MC samples, where the largest inefficiency in the
fiducial region is $<5\%$. However, since this inefficiency is identical for the \znzn\ and
\znxn\ samples, it has no impact on the ratio, so no correction is applied. An additional
correction to the overall normalization results from independent \PbPb\ EMD scatterings which
occur in the same crossing as an event of interest (EMD pile-up). Corrections are derived using
the methodology from Refs.~\cite{HION-2016-02,HION-2021-16}, and the resulting event-averaged
correction to the normalization is $13(7)\%$ for the \znzn(\znxn) topology.

\begin{figure}[!tb]
\centerline{
\subfloat[]{\includegraphics[width=0.470\textwidth]{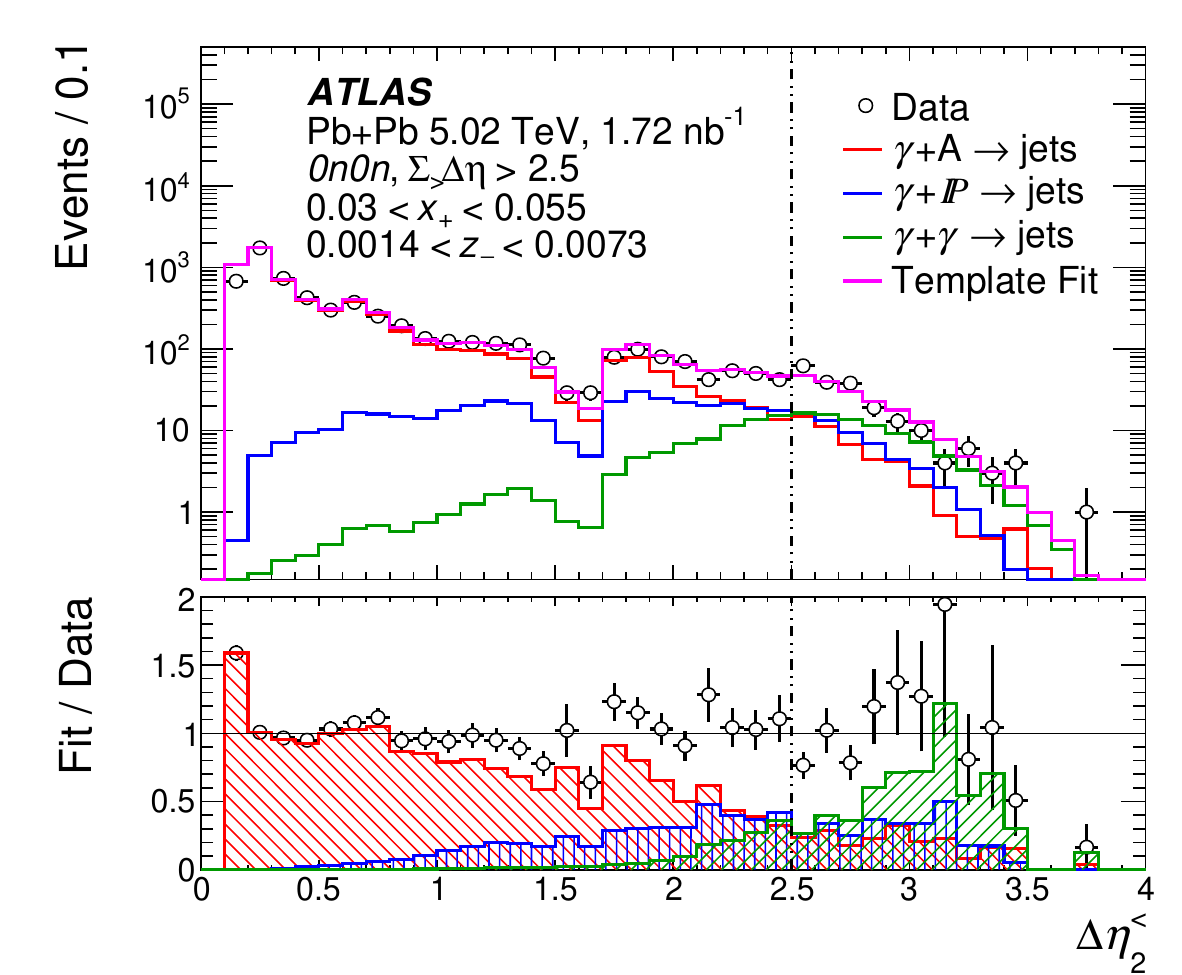}}
\subfloat[]{\includegraphics[width=0.530\textwidth]{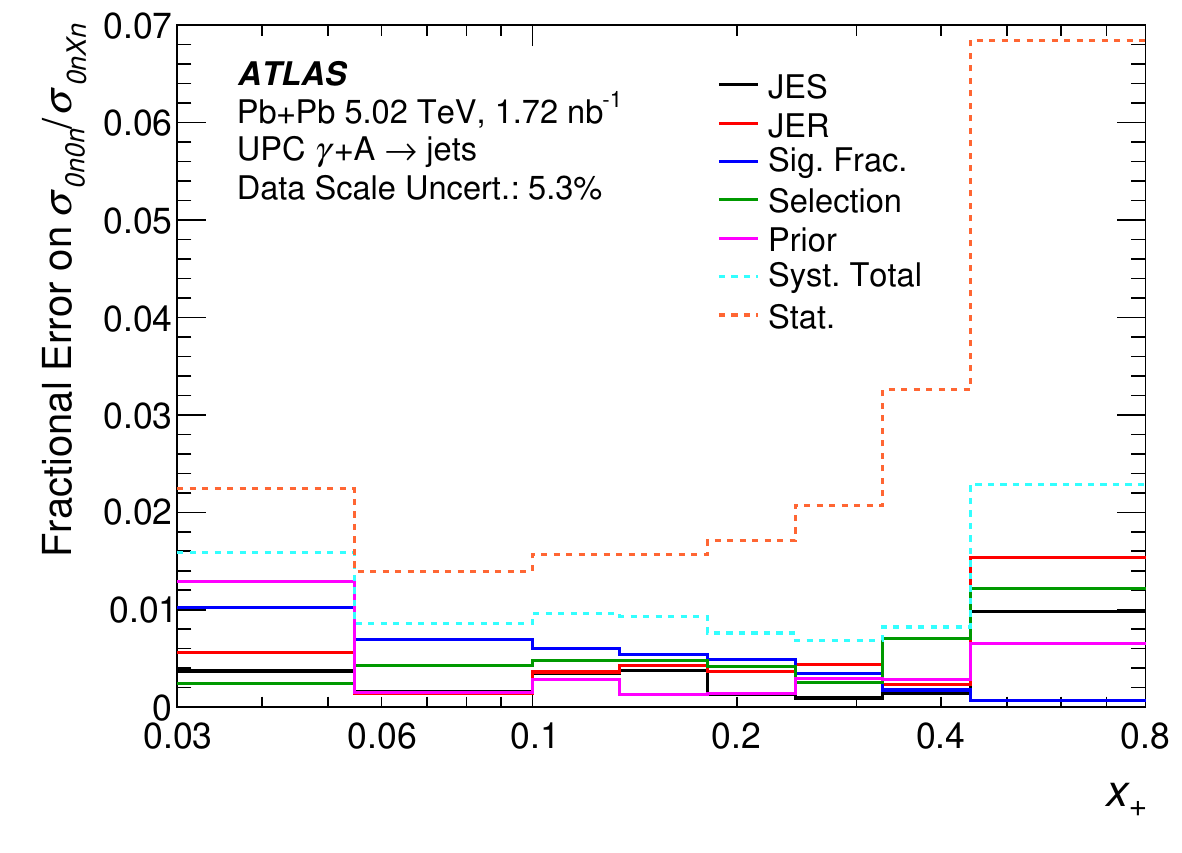}}
}
\caption{
(a) An example template fit to the distribution of \edgeGapTwoMin, the smallest gap from an
edge of the ATLAS detector to the second-closest ID track or topo-cluster, for \znzn\
UPC jet events in a kinematic interval with substantial contributions from all
three components. The \gamAjets\ template is derived from \znxn\ data, while
the \gamPomJets\ and \gamGamJets\ templates are derived from \pythia\ MC.
The lower panel shows the ratio of the different process templates
with respect to data, and the markers show the ratio of the full template fit to
the data. The dashed line indicates the upper selection on \edgeGapTwoMin\ applied in
order to select a higher-purity \gamAjets\ sample.
(b) The fractional uncertainties on the \znzn/\znxn\ cross-section ratio from different
sources of systematic uncertainty which contribute to this measurement as a function
of \xpos.
The fractional statistical uncertainty is also shown, and it is larger than the combined
bin-to-bin systematic uncertainty in all bins. The fully correlated scale uncertainty of
$\pm 5.3\%$ is composed of $\pm0.1\%$ (JES+JER) $\pm2.5\%$ (Prior) $\pm0.07\%$ (Sig. Frac.)
$\pm 0.6\%$ (EMD pile-up) $\pm 4.7\%$ (EMD).
}
\label{fig:SystBreakdown}
\end{figure}

While \sumGapMax characterizes the photon-going direction, a different rapidity
gap variable, \edgeGapTwo, is more useful to describe the other side of the collision.
The \edgeGapTwo variable is defined as the rapidity interval between
the edge of the ATLAS detector and the second-closest ID track or topological
cluster, on a given side of the detector. The second-closest track or topo-cluster is
chosen because two topo-clusters originating from calorimeter noise rarely occur
nearby in the same event, while physical deposits are rarely alone, greatly reducing
the sensitivity to detector mis-reconstruction effects. The minimum \edgeGapTwo\ between
the two detector sides, \edgeGapTwoMin, yields distinct distributions for
all three processes which contribute to the total \znzn\ cross-section: \gamA,
\gamPom, and \gamgam. The fraction of the total cross-section associated with
\gamA\ processes is assessed via template fits to
the \edgeGapTwoMin\ distribution that is explained in greater
detail in Ref.~\cite{OtherPaperPRC}. The fits are performed in intervals of $(\zneg, \xpos)$ that match
the binning of the measured cross-section. An example result from the template-fitting
procedure is shown in panel (a) of Figure~\ref{fig:SystBreakdown}. The
measured gap distribution is described as the sum of templates corresponding
to the three different processes which contribute to \znzn\ UPC jet production. The
shape at $\edgeGapTwoMin = 1.7$ results from the structure of the ATLAS calorimeter
which affects the topo-cluster reconstruction. An additional rapidity-gap selection is also applied,
requiring $\edgeGapTwoMin < 2.5$, in order to reduce sensitivity to uncertainty in the
template fitting procedure. In the measured kinematic region, only the
lowest two \xpos\ intervals have a significant background fraction
($5\%$ and $1\%$, respectively), and the higher-\xpos\ intervals have
negligible contributions from \gamPom\ or \gamgam\ processes.

Once the data are corrected for inefficiencies in the event selections and for residual
backgrounds, the fully corrected yields for \gamAjets\ production as a
function of \HT, \xpos, and \zneg\ are obtained. These distributions are binned
differentially in \xpos\ using two logarithmic bins between $0.03$ and $0.1$ and
seven logarithmic bins from $0.1$ to $0.8$, where the two highest-\xpos\ bins
are merged due to limited statistical precision. The binned yields are corrected for
residual detector effects with a Bayesian unfolding procedure~\cite{DAgostini:1994fjx}
using the \verb+RooUnfold+ package~\cite{Adye:2011gm} with two iterations.
The unfolding response matrices are created from the \gamAjets\ \pythia\ MC
sample, which is re-weighted so its reconstructed distribution in \HT, \xpos, and \zneg\ matches
that of the data distribution it is used to unfold (\znzn\ or \znxn).
The \znzn\ and \znxn\ yields are separately unfolded three-dimensionally in \HT, \xpos, and \zneg,
and the resulting differential cross-sections are integrated over \HT.
Statistical uncertainties on the unfolded data distributions are computed using
$1000$ stochastic variations on the data and response
matrices~\cite{ATL-PHYS-PUB-2021-011}. The effect of the unfolding
on the \znzn/\znxn\ cross-section ratio is typically smaller than the
statistical uncertainty.

Systematic uncertainties on the measured cross-section ratios are assessed to account
for effects arising from the
event selections, jet energy measurement, sensitivity to the unfolding prior, and
signal fraction determination. The first three sources partially cancel for
the \znzn/\znxn\ ratio, but the signal fraction
uncertainty only impacts the \znzn\ sample and thus does not exhibit any cancellation.
The event selection uncertainties are assessed for the gap requirement
($\sumGapMax > 2.5$) by tightening it to $\sumGapMax > 3.0$ and re-computing both the cross-sections
and gap selection efficiency corrections with this new requirement.
The jet response systematic uncertainties are propagated from the uncertainties derived in
Ref.~\cite{HION-2022-15}, following the standard ATLAS jet calibration procedure for
Run 2~\cite{JETM-2018-05}. The jet response is varied when constructing the response
matrices to account for uncertainty in the jet energy scale (JES) and resolution (JER), and the results
unfolded with the varied response are compared to results using the nominal jet calibration.
An additional \pT-dependent uncertainty is assigned on the jet energy scale to
account for extrapolating the jet energy measurement from $15$~\GeV, its
nominal limit of applicability, to $13$~\GeV.
The residual sensitivity to the unfolding prior uncertainty is determined by
modifying the re-weighting prior distribution in the response matrices. For each variation,
the nominal result is compared to
a result where both distributions are unfolded with the same prior, either \znzn\ or \znxn.

The numerator of the \znzn/\znxn\ cross-section ratio is sensitive to the peripheral
\gamA\ signal fractions obtained via the template-fitting procedure. The associated
systematic uncertainty is evaluated through three variations to the extraction of the
signal fraction. The fit range in \edgeGapTwoMin\ is varied,
where the first variation removes two additional bins ($0.2$ units in $\eta$) from
the start of the fit and the second variation removes the five last bins ($0.5$ units in $\eta$). Uncertainties
are also applied to account for imperfect modeling of the topo-cluster reconstruction
in simulation. An additional variation is applied where the proportion of direct and resolved photon
events in the $\gamma+\gamma$ template is varied to match the fractions observed in Ref.~\cite{HION-2022-15}.
Additionally, an uncertainty is assessed to account for potential mis-modeling of the shape of the \gamPom\
contribution by performing the fits using modified MC \gamPom\ samples
corresponding to conservative variations in the size or shape of the Pb nucleus. These samples
were obtained by performing re-weightings of the nominal sample as a function of the
momentum-transfer of the pomeron. The varied \gamAjets\
signal fractions are used to re-derive cross-sections, which are
then unfolded and compared to the nominal distribution to derive the systematic
uncertainties. For more details on this procedure, see Ref.~\cite{OtherPaperPRC}.
While this uncertainty is typically quite small in the measured
region, the uncertainty in the unfolded results is slightly larger due to
possible migration of backgrounds from outside the fiducial region.

A summary of the different contributions to the experimental uncertainty on
the \znzn/\znxn\ cross-section ratio is shown in the right panel of Figure~\ref{fig:SystBreakdown}.
The statistical uncertainty is dominant for all bins, but the largest sources
of systematic uncertainty arise from the signal fraction and unfolding prior
sensitivity at low \xpos.
Additionally, fully correlated uncertainties arise
due to the luminosity measurement and correction for EMD pile-up. The luminosity
uncertainty for the full 2018 Pb+Pb dataset is $2.0\%$, derived using methods
described in Ref. \cite{DAPR-2021-01}, using the LUCID-2 detector for the baseline luminosity
measurements \cite{LUCID2}. While this uncertainty impacts the absolutely normalized cross-sections,
it has no impact on the \znzn/\znxn\ cross-section ratios. The uncertainty on the absolute
cross-section arising due to the correction for EMD pile-up is obtained by varying the \PbPb\
dissociative cross-section within its uncertainties~\cite{PhysRevLett.109.252302}. This variation
yields a fractional uncertainty of $0.6\%$ on the absolute cross-sections, with a
negligible impact on ratios. An additional uncertainty, which is found to be fully correlated
between bins, is applied to account for uncertainty on the correction for additional EMD
within the same event, as measured in Refs.~\cite{OtherPaperPRC} and ~\cite{HION-2022-15}
for the \znzn\ and \znxn cases, respectively.


%

%
%

%
%
%

\begin{figure}[!t]
\centerline{
\includegraphics[width=0.49\textwidth]{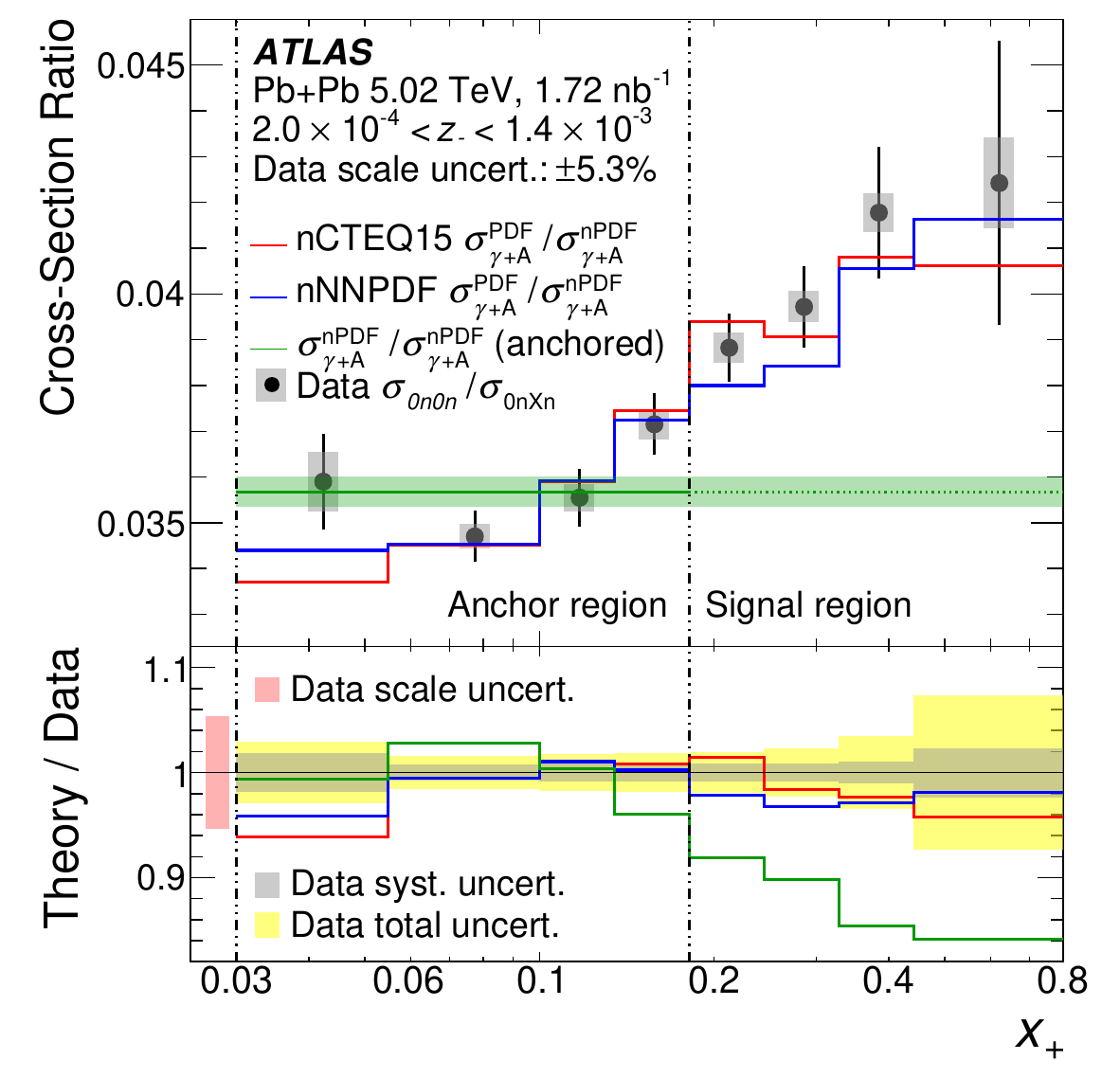}
}
\caption{
The ratio of \gamAjets\ cross-sections for the \znzn\ neutron topology (numerator)
and \znxn\ neutron topology (denominator) as a function of \xpos. Systematic uncertainties are shown as shaded
bands, while vertical lines represent combined systematic and statistical uncertainties. The
fully correlated component of the systematic uncertainty is subtracted from the
bin-to-bin contributions and denoted by the $5.3\%$ scale uncertainty.
Theoretical predictions are shown for the
ratio of the \gamAjets\ cross-section computed in \pythia\ with unmodified PDFs
divided by the result with modified nPDFs, for nCTEQ~\cite{Duwentaster:2021ioo}
(red) and nNNPDF~\cite{AbdulKhalek:2022fyi} (blue). The green
line represents the null hypothesis that the \znzn\ and \znxn\ samples
exhibit the same nPDF modification. Each prediction is scaled to match
the integral in the anchor region, and the uncertainty on this normalization
is shown as the shaded green band.
The bottom panel shows the ratio of the different predictions to the data,
with statistical and total uncertainties shown as shaded yellow and gray bands,
respectively.
}
\label{fig:ResultsNPDF}
\end{figure}

The fully unfolded ratio of \gamAjets\ cross-sections in \znzn\ and \znxn\ topologies
as a function of \xpos\ is shown in Figure~\ref{fig:ResultsNPDF}. For these comparisons,
both sets of unfolded cross-sections are corrected for additional EMD within the same
event~\cite{OtherPaperPRC,HION-2022-15}.
Since it is possible for differences between the shapes of the \znzn\ and \znxn\ cross-sections
to arise from modifications to the photon flux probed by each class of events,
results are restricted to the range $2\times10^{-4} < \zneg < 1.4\times10^{-3}$. This
interval was chosen to select a sample with sufficiently small \zneg\ such that,
according to calculations following a
procedure from Ref.~\cite{Eskola:2024fhf}, the results are insensitive to possible
modifications in the photon flux in the \znzn\ sample. This selection removes about
$48.4\%$($51.8\%$) of the \znzn(\znxn) sample.
These data show a clearly increasing trend from small to large \xpos,
indicating a modified slope in the \znzn\ cross-section relative to the \znxn case.
The data are compared with theoretical predictions that assume the \znzn\ selection
corresponds to collisions with unmodified free nucleons.
The predicted cross-sections with nPDF modifications
($\sigma^\mathrm{nPDF}_{\gamA}$) for these comparisons are derived from
nCTEQ15 WZ+SIH~\cite{Duwentaster:2021ioo} (red) and nNNPDF
3.0~\cite{AbdulKhalek:2022fyi} (blue). The corresponding cross-sections computed with baseline free
nucleon PDFs ($\sigma^\mathrm{PDF}_{\gamA}$) are derived from
CT18~\cite{Hou:2019efy} and NNPDF 3.1~\cite{Ball:2022oua},
respectively. All samples account for the isospin of lead. As the theoretical
predictions provide no absolute normalization for the ratio,
they are scaled to match the ratio of the cross-section integrals in the region
$(0.03 < \xpos < 0.1811)$, the choice of which is motivated below. The agreement in shape between
the theoretical comparisons and the data is consistent with photonuclear interactions at the
nuclear periphery behaving as if the nucleons are free and unmodified by nuclear effects.

To evaluate the statistical significance of the observed modification
to the \xpos\ distribution, a test is performed that takes
advantage of the fact that the impact of the nuclear modifications
on the measured ratio is monotonic in \xpos. Namely, the
measured \xpos\ range is divided into ``anchor'' and ``signal'' regions
within which the \znzn\ and \znxn\ cross-sections are separately summed.
The \znzn/\znxn\ cross-section ratios are then computed in each region, and the ``null''
hypothesis that the ratios are the same in the two regions is tested.
The division between signal and anchor regions was determined prior to
examining the data, using purely MC samples. Specifically, hypothetical \znzn\
cross-sections were computed using \pythia \gamAjets\ samples with different combinations of modified
and unmodified nPDFs, which were re-sampled to match the number of events in the data.
Ratios to simulated \znxn\ cross-sections were computed in both the anchor
and signal regions using different boundaries in the measured \xpos range. Taking into
account statistical and systematic uncertainties on the data, these studies
indicated an optimal division with anchor and signal regions of $(0.03
< \xpos < 0.1811)$ and $(0.1811 < \xpos < 0.8)$, respectively.
An expected significance of $6.5~\sigma$ is obtained in MC studies
where the \znzn\ cross-section is computed from free
nucleon -- {\em i.e. unmodified} -- PDFs and the \znxn\ using
nCTEQ~\cite{Duwentaster:2021ioo} nPDFs. Then, computing the significance in
data using this division between anchor
and signal regions, the measured cross-section ratios exclude
the null hypothesis with an observed significance
of $6.0\sigma$, which is mostly insensitive to variations in the boundary
between the anchor and signal regions. This result indicates an observation
of a significant difference in the nuclear PDFs probed by the
\znzn\ and \znxn\ selections within the signal \xpos range and,
thus, of a significant $b_\text{A}$-dependence in the nPDF.%

\FloatBarrier


%

%
%
%

%
%
%

In summary, ATLAS has studied jet photoproduction
in ultra-peripheral 5.02~\TeV \PbPb collisions in which both
nuclei remain intact, comparing them to collisions where one nucleus dissociates . The lack of nuclear
breakup is determined by the absence of beam-rapidity neutrons in
ZDCs. At small \zneg (larger $|\yjets|$), the non-diffractive \gamA cross-section dominates, and a
measurement of jet photoproduction in hard \gamA scattering with no
nuclear breakup is possible.
The ratio of jet cross-sections between the \znzn\ and \znxn\ collision events
is measured, and it shows a clear dependence on \xpos. This feature indicates
a different, \xpos-dependent, modification of the nPDFs in the \znzn\ and \znxn\
topologies. Since the \znzn selection is expected to preferentially select peripheral
(large $b_\text{A}$) collisions while the \znxn requirements selects
all other \gamA collisions, the observed difference is consistent with
a lack of nuclear modifications in  peripheral events. Such a
conclusion could have consequences for interpreting
results from the heavy ion programs at RHIC, the LHC, and the
HL-LHC. It also demonstrates a method for studying the
$b_\text{A}$-dependence of nPDF modifications at both the HL-LHC and the future electron-ion collider.
The observed significance in rejecting the null hypothesis that the \znzn
collisions probe the same nuclear PDFs as \znxn collisions
is $6.0\sigma$, providing the first evidence that the parton
distributions of nucleons near the edge of a nucleus differ from those of nucleons
near its center.


%

%
%
%
%
%
\section*{Acknowledgements}
%

%
%
%
%

%
%

%
%

We thank CERN for the very successful operation of the LHC and its injectors, as well as the support staff at
CERN and at our institutions worldwide without whom ATLAS could not be operated efficiently.

The crucial computing support from all WLCG partners is acknowledged gratefully, in particular from CERN, the ATLAS Tier-1 facilities at TRIUMF/SFU (Canada), NDGF (Denmark, Norway, Sweden), CC-IN2P3 (France), KIT/GridKA (Germany), INFN-CNAF (Italy), NL-T1 (Netherlands), PIC (Spain), RAL (UK) and BNL (USA), the Tier-2 facilities worldwide and large non-WLCG resource providers. Major contributors of computing resources are listed in Ref.~\cite{ATL-SOFT-PUB-2026-001}.

We gratefully acknowledge the support of ANPCyT, Argentina; YerPhI, Armenia; ARC, Australia; BMWFW and FWF, Austria; ANAS, Azerbaijan; CNPq and FAPESP, Brazil; NSERC, NRC and CFI, Canada; CERN; ANID, Chile; CAS, MOST and NSFC, China; Minciencias, Colombia; MEYS CR, Czech Republic; DNRF and DNSRC, Denmark; IN2P3-CNRS and CEA-DRF/IRFU, France; SRNSFG, Georgia; BMFTR, HGF and MPG, Germany; GSRI, Greece; RGC and Hong Kong SAR, China; ICHEP and Academy of Sciences and Humanities, Israel; INFN, Italy; MEXT and JSPS, Japan; CNRST, Morocco; NWO, Netherlands; RCN, Norway; MNiSW, Poland; FCT, Portugal; MNE/IFA, Romania; MSTDI, Serbia; MSSR, Slovakia; ARIS and MVZI, Slovenia; DSI/NRF, South Africa; MICIU/AEI, Spain; SRC and Wallenberg Foundation, Sweden; SERI, SNSF and Cantons of Bern and Geneva, Switzerland; NSTC, Taipei; TENMAK, T\"urkiye; STFC/UKRI, United Kingdom; DOE and NSF, United States of America.

Individual groups and members have received support from BCKDF, CANARIE, CRC and DRAC, Canada; CERN-CZ, FORTE and PRIMUS, Czech Republic; COST, ERC, ERDF, Horizon 2020, ICSC-NextGenerationEU and Marie Sk{\l}odowska-Curie Actions, European Union; Investissements d'Avenir Labex, Investissements d'Avenir Idex and ANR, France; DFG and AvH Foundation, Germany; Herakleitos, Thales and Aristeia programmes co-financed by EU-ESF and the Greek NSRF, Greece; BSF-NSF and MINERVA, Israel; NCN and NAWA, Poland; La Caixa Banking Foundation, CERCA and AGAUR programs from Generalitat de Catalunya and PROMETEO and GenT Programmes Generalitat Valenciana, Spain; G\"{o}ran Gustafssons Stiftelse, Sweden; The Royal Society and Leverhulme Trust, United Kingdom; Eric and Wendy Schmidt Fund for Strategic Innovation, United States of America.

In addition, individual members wish to acknowledge support from CERN: European Organization for Nuclear Research (CERN DOCT); Chile: Agencia Nacional de Investigaci\'on y Desarrollo (ANID FONDECYT reg. 1230987, FONDECYT 1230812, FONDECYT 1240864, Fondecyt 3240661, Fondecyt Regular 1240721); China: Chinese Ministry of Science and Technology (MOST-2023YFA1605700, MOST-2023YFA1609300), National Natural Science Foundation of China (NSFC - 12175119, NSFC 12275265); Czech Republic: Czech Science Foundation (GACR - 24-11373S), Ministry of Education Youth and Sports (ERC-CZ-LL2327, FORTE CZ.02.01.01/00/22\_008/0004632), PRIMUS Research Programme (PRIMUS/21/SCI/017); EU: H2020 European Research Council (ERC - 101002463); European Union: European Research Council (BARD No. 101116429, ERC - 948254, ERC 101089007), European Regional Development Fund (HE COFUND GA No.101081355, ERDF), European Union, Future Artificial Intelligence Research (FAIR-NextGenerationEU PE00000013), Italian Center for High Performance Computing, Big Data and Quantum Computing (ICSC, NextGenerationEU), Marie Sklodowska-Curie Actions (ERC Horizon-2020, GA 956086, GAP-101168829); France: Agence Nationale de la Recherche (ANR-21-CE31-0013, ANR-21-CE31-0022, ANR-22-EDIR-0002, ANR-24-CE31-0504-01); Germany: Deutsche Forschungsgemeinschaft (DFG - 469666862, DFG - CR 312/5-2); China: Research Grants Council (GRF); Italy: Istituto Nazionale di Fisica Nucleare (ICSC, NextGenerationEU), Ministero dell'Università e della Ricerca (NextGenEU 153D23001490006 M4C2.1.1, NextGenEU I53D23000820006 M4C2.1.1, NextGenEU I53D23001490006 M4C2.1.1, SOE2024\_0000023); Japan: Japan Society for the Promotion of Science (JSPS KAKENHI  JP25H0063, JSPS KAKENHI JP22H01227, JSPS KAKENHI JP22H04944, JSPS KAKENHI JP22KK0227, JSPS KAKENHI JP24K23939, JSPS KAKENHI JP24KK0251, JSPS KAKENHI JP25H00650, JSPS KAKENHI JP25H01291, JSPS KAKENHI JP25K01011, JSPS KAKENHI JP25K01023); Norway: Research Council of Norway (RCN-314472); Poland: Ministry of Science and Higher Education (IDUB AGH, POB8, D4 no 9722), Polish National Science Centre (NCN 2021/42/E/ST2/00350, NCN OPUS 2023/51/B/ST2/02507, NCN OPUS nr 2022/47/B/ST2/03059, NCN UMO-2019/34/E/ST2/00393, UMO-2022/47/O/ST2/00148, UMO-2023/49/B/ST2/04085, UMO-2023/51/B/ST2/00920, UMO-2024/53/N/ST2/00869); Portugal: Foundation for Science and Technology (FCT); Spain: Agència de Gestió d'Ajuts Universitaris i de Recerca. (AGAUR - 2023 BP 00141), Generalitat Valenciana (ASFAE/2022/008), Ministry of Science and Innovation (MCIN \& NextGenEU PCI2022-135018-2, MICIN \& FEDER PID2021-125273NB, RYC2019-028510-I, RYC2020-030254-I, RYC2021-031273-I, RYC2022-038164-I), Ministerio de Ciencia, Innovación y Universidades/Agencia Estatal de Investigaci\'on (EU NextGenerationEU (PRTR-C17.I1), PID2022-142604OB-C22); Sweden: Carl Trygger Foundation (Carl Trygger Foundation CTS 22:2312), Swedish Research Council (Swedish Research Council 2023-04654, VR 2021-03651, VR 2022-03845, VR 2022-04683, VR 2023-03403, VR 2024-05451), Knut and Alice Wallenberg Foundation (KAW 2018.0458, KAW 2022.0358, KAW 2023.0366); Switzerland: Swiss National Science Foundation (SNSF - PCEFP2\_194658); United Kingdom: Royal Society (NIF-R1-231091); United States of America: U.S. Department of Energy (ECA DE-AC02-76SF00515), Neubauer Family Foundation.

%
%


%
\clearpage
\appendix
\part*{End matter}
\FloatBarrier
\addcontentsline{toc}{part}{End matter}

\begin{figure}[!b]
\centerline{
\includegraphics[width=1.0\textwidth]{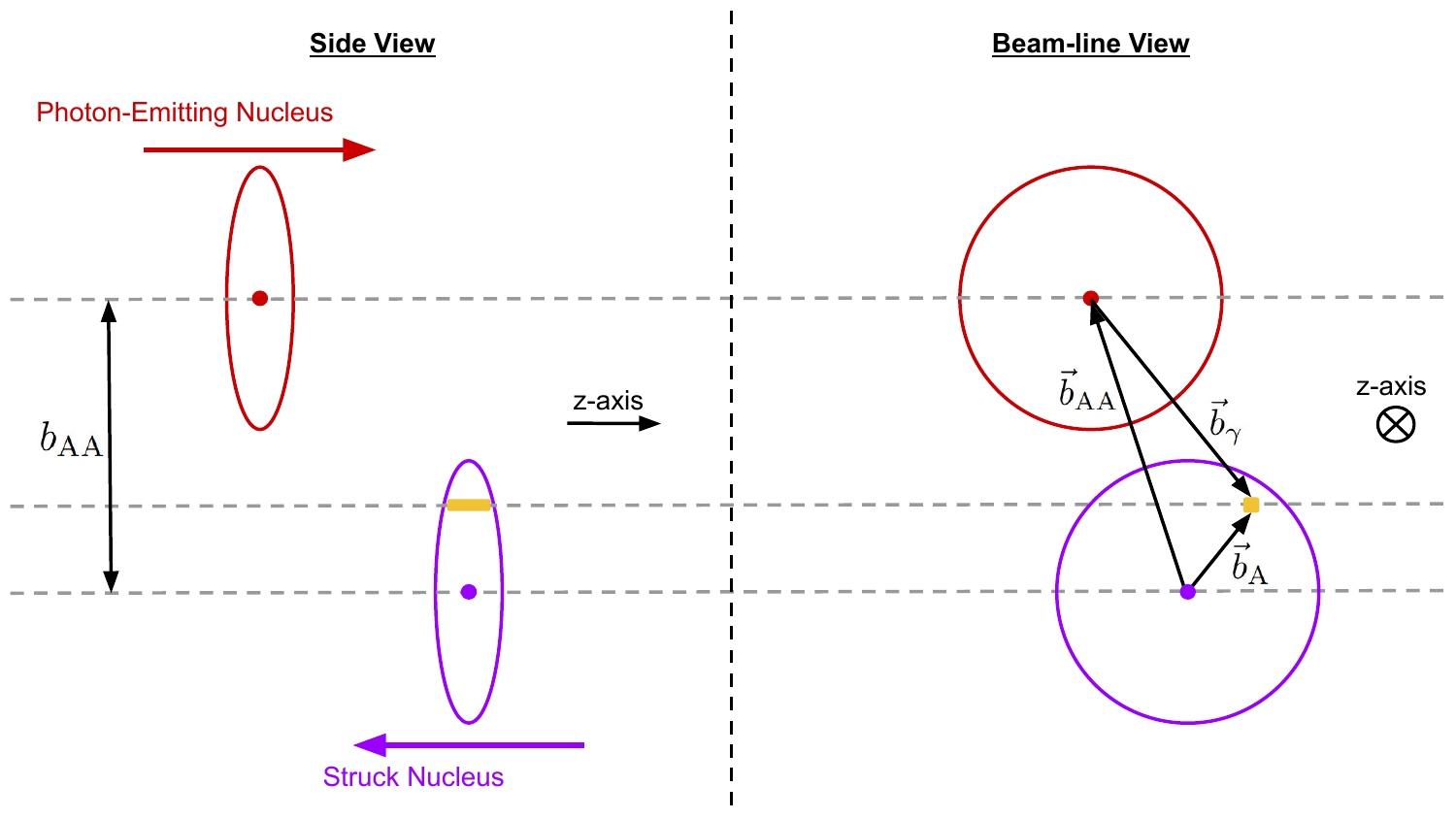}
}
\caption{
A diagram of an ultra-peripheral photonuclear collision between
two heavy ions, shown from a side view (left) or beam-line view (right).
The photon-emitting nucleus (red) and struck nucleus
(purple) are separated by impact parameter $\vec{b}_\mathrm{AA}$.
The two vectors $\vec{b}_\gamma$ and $\vec{b}_\mathrm{A}$
indicate the distance from the site of the \gamA\ interaction (orange)
to the centers of the photon-emitting and struck nuclei, respectively.
}
\label{fig:ImpactParam}
\end{figure}

Figure~\ref{fig:ImpactParam} demonstrates the geometry of a photonuclear
collision, where the two colliding nuclei are separated by the
two-nucleus impact parameter ($\vec{b}_\mathrm{AA}$) which satisfies
$|\vec{b}_\mathrm{AA}| > 2R_\mathrm{A}$. The impact parameters of the
collision with respect to the two nuclei are $\vec{b}_\gamma$ and
$\vec{b}_\mathrm{A}$ for the photon-emitting and struck nuclei,
respectively. These impact parameters satisfy
$\vec{b}_\mathrm{AA} = \vec{b}_\mathrm{A} - \vec{b}_\gamma$.

\begin{figure}[!b]
\centerline{
\includegraphics[width=0.7\textwidth]{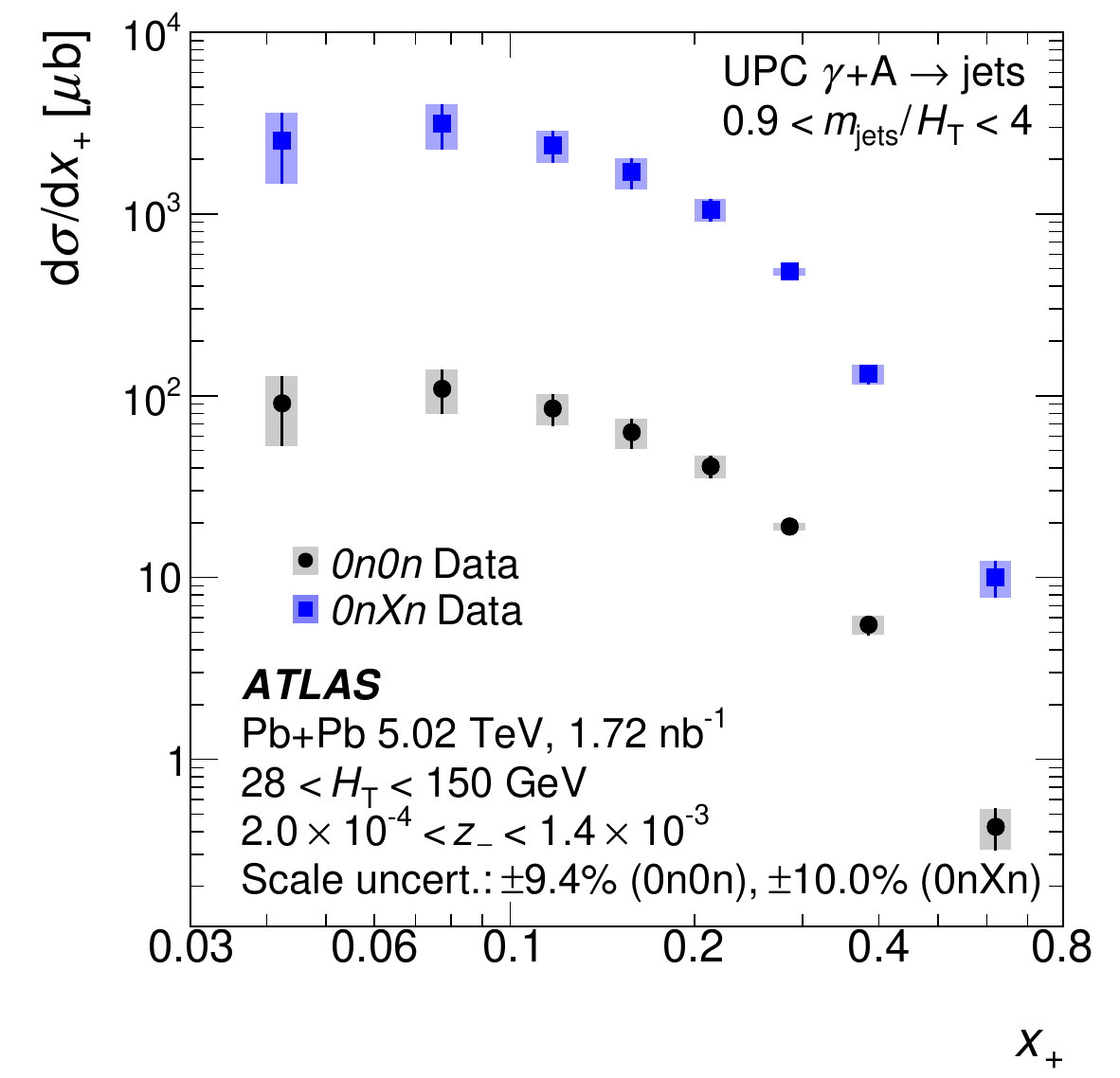}
}
\caption{
Cross-sections for photonuclear jet production measured with \znzn\ (black)
and \znxn\ (blue) neutron topologies. Systematic uncertainties are shown as
shaded bands, and the combined statistical and systematic uncertainties are
shown as vertical lines. The
fully correlated component of the systematic uncertainty is subtracted from the
bin-to-bin contributions and denoted by the \znzn\ and \znxn\ scale uncertainties.
The kinematic selections applied here are identical to those used in
Figure~\ref{fig:ResultsNPDF}.
}
\label{fig:CrossSecGamA}
\end{figure}

The cross-sections for \gamAjets\ production in \znzn\ and \znxn\ neutron
topologies are shown in Figure~\ref{fig:CrossSecGamA}. These results provide
the numerator and denominator, respectively, for the
ratio shown in Figure~\ref{fig:ResultsNPDF}. The cross-sections are corrected
for the impact of additional EM dissociation, as measured in
Ref.~\cite{OtherPaperPRC}. In each case, the data scale uncertainties and
point-to-point systematic uncertainties are dominated by uncertainty on the
jet energy scale and resolution. The similar shapes of the cross-sections,
with the \znzn\ cross-section about $0.04$ times the \znxn,
is consistent with the \znzn\ measuring a small fraction of the total
non-diffractive \gamA\ cross-section where no forward neutrons are emitted
by either nucleus.

%
%
%
%
%
%
%
%
%
%
%
%
%
%
%

%
%
%
%
%
%
%
%
%
%


\FloatBarrier
\clearpage
\printbibliography
\clearpage
 
\begin{flushleft}
\hypersetup{urlcolor=black}
{\Large The ATLAS Collaboration}

\bigskip

\AtlasOrcid[0000-0002-6665-4934]{G.~Aad}$^\textrm{\scriptsize 103}$,
\AtlasOrcid[0000-0001-7616-1554]{E.~Aakvaag}$^\textrm{\scriptsize 17}$,
\AtlasOrcid[0000-0002-5888-2734]{B.~Abbott}$^\textrm{\scriptsize 121}$,
\AtlasOrcid[0000-0002-0287-5869]{S.~Abdelhameed}$^\textrm{\scriptsize 84b}$,
\AtlasOrcid[0000-0002-1002-1652]{K.~Abeling}$^\textrm{\scriptsize 55}$,
\AtlasOrcid[0000-0001-5763-2760]{N.J.~Abicht}$^\textrm{\scriptsize 49}$,
\AtlasOrcid[0000-0002-8496-9294]{S.H.~Abidi}$^\textrm{\scriptsize 30}$,
\AtlasOrcid[0009-0003-6578-220X]{M.~Aboelela}$^\textrm{\scriptsize 45}$,
\AtlasOrcid[0000-0002-9987-2292]{A.~Aboulhorma}$^\textrm{\scriptsize 36e}$,
\AtlasOrcid[0000-0001-5329-6640]{H.~Abramowicz}$^\textrm{\scriptsize 155}$
% ...
\bigskip
\\

% ...
$^{17}$Department for Physics and Technology, University of Bergen, Bergen; Norway.\\
% ...
$^{30}$Physics Department, Brookhaven National Laboratory, Upton NY; United States of America.\\
% ...
$^{36}$$^{(a)}$Facult\'e des Sciences Ain Chock, Universit\'e Hassan II de Casablanca;$^{(b)}$Facult\'{e} des Sciences, Universit\'{e} Ibn-Tofail, K\'{e}nitra;$^{(c)}$Facult\'e des Sciences Semlalia, Universit\'e Cadi Ayyad, LPHEA-Marrakech;$^{(d)}$LPMR, Facult\'e des Sciences, Universit\'e Mohamed Premier, Oujda;$^{(e)}$Facult\'e des sciences, Universit\'e Mohammed V, Rabat;$^{(f)}$Institute of Applied Physics, Mohammed VI Polytechnic University, Ben Guerir; Morocco.\\
% ...
$^{45}$Physics Department, Southern Methodist University, Dallas TX; United States of America.\\
% ...
$^{49}$Fakult\"{a}t Physik, Technische Universit{\"a}t Dortmund, Dortmund; Germany.\\
% ...
$^{55}$II. Physikalisches Institut, Georg-August-Universit\"{a}t G\"ottingen, G\"ottingen; Germany.\\
% ...
$^{84}$$^{(a)}$Khalifa University of Science and Technology, Abu Dhabi;$^{(b)}$New York University Abu Dhabi, Abu Dhabi;$^{(c)}$United Arab Emirates University, Al Ain;$^{(d)}$University of Sharjah, Sharjah; United Arab Emirates.\\
% ...
$^{103}$CPPM, Aix-Marseille Universit\'e, CNRS/IN2P3, Marseille; France.\\
% ...
$^{121}$Homer L. Dodge Department of Physics and Astronomy, University of Oklahoma, Norman OK; United States of America.\\
% ...
$^{155}$Raymond and Beverly Sackler School of Physics and Astronomy, Tel Aviv University, Tel Aviv; Israel.\\
% ...
$^{a}$ Also at Affiliated with an institute formerly covered by a cooperation agreement with CERN.\\
$^{b}$ Also at An-Najah National University, Nablus; Palestine.\\
$^{c}$ Also at Borough of Manhattan Community College, City University of New York, New York NY; United States of America.\\
$^{d}$ Also at Center for Interdisciplinary Research and Innovation (CIRI-AUTH), Thessaloniki; Greece.\\
$^{e}$ Also at Centre of Physics of the Universities of Minho and Porto (CF-UM-UP); Portugal.\\
$^{f}$ Also at CERN, Geneva; Switzerland.\\
% ...

\end{flushleft}


%

%
%
%
%

%
%
%
%

\end{document}